\documentclass[preprintnumbers]{revtex4}

\usepackage{graphicx}
\def\pairof#1{#1^+ #1^-}
\def\ee{\pairof{e}}
\def\ww{W^+W^-}
\def\ra{\rightarrow}
\def\wwl{W_{\rm L}W_{\rm L}}
\def\dkg{\Delta\kappa_\gamma}

\begin{document}
\bibliographystyle{apsrev}

\def\lsim{\mathrel{\raise.3ex\hbox{$<$\kern-.75em\lower1ex\hbox{$\sim$}}}}
\def\gsim{\mathrel{\raise.3ex\hbox{$>$\kern-.75em\lower1ex\hbox{$\sim$}}}}
\renewcommand{\O}{0}


\preprint{P1-WG4}
\preprint{P3-50}

\title[]{Electroweak Symmetry Breaking by Strong Dynamics\\
and the Collider Phenomenology\footnote{This is the report 
of the Strong Electroweak Symmetry Breaking Working Group 
at 2001 Snowmass summer studies.
}}


\author{ {\bf Conveners:} Timothy L. Barklow, R. Sekhar Chivukula, 
Joel Goldstein, Tao Han}
\email[]{timb@slac.stanford.edu}
\email[]{joelg@fnal.gov}
\email[]{sekhar@bu.edu}
\email[]{than@pheno.physics.wisc.edu}
\affiliation{Stanford Linear Accelerator Center,
Stanford University, Stanford, California 94309 USA}
\affiliation{Fermilab}
\affiliation{Department of Physics, Boston University}
\affiliation{Department of Physics, University of Wisconsin, 
Madison, WI 53706}

\author{{\bf Working group members:}
D. Dominici, S. Godfrey, R. Harris, H.-J. He, P. Kalyniak, 
W. Kilian, K. Lynch, T. Liu, J. Mizukoshi, M. Narain,
T. Ohl, M. Popovic, A. Skuja, T. Tait, W. Walkowiak,
J. Womersley}
\noaffiliation

\date{\today}

\begin{abstract}
We discuss the possible signatures in the electroweak
symmetry breaking sector by new strong dynamics at 
future hadron colliders such as the Tevatron upgrade,
the LHC and VLHC, and $e^+e^-$ linear colliders. Examples
include a heavy Higgs-like scalar resonance, a heavy 
Technicolor-like vector resonance and pseudo-Goldstone
states, non-resonance signatures via enhanced gauge-boson 
scattering and fermion compositeness.
\end{abstract}

\maketitle

\medskip
\section{Introduction}

Particle physics is on the verge of major discovery. General
arguments indicate that new physics in the electroweak symmetry
breaking sector must show up below the scale of 1 TeV. The 
experiments at the Tevatron and next generation high energy 
colliders such as the LHC and a TeV $e^+e^-$ linear collider
will fully explore the new physics at the electroweak scale.

In theories of dynamical electroweak symmetry breaking, the electroweak
interactions are broken to electromagnetism by the vacuum expectation
value of a fermion bilinear. These theories may thereby avoid the
introduction of fundamental scalar particles, of which we have no
examples in nature thus far. Prominent examples include Technicolor, 
topcolor, and related models. If the new dynamical scale is somewhat
higher than 1 TeV, then the low energy effects or the early 
signature at collider experiments may be 
anomalous gauge boson interactions, enhanced $WW$ scattering signals,
or contact 4-fermion interactions. In this report, we first
briefly introduce the dynamical electroweak symmetry breaking
models and parameterization of the anomalous couplings. 
We then summarize the collider sensitivities to probe
the new dynamics at future $e^+e^-$ linear colliders in Sec.~II, 
and at hadron colliders in Sec.~III.

\subsection{Technicolor}

The earliest models\cite{Weinberg,Susskind} of dynamical electroweak
symmetry breaking\cite{Chivukula:1998if,Simmons:2001zt} include a new
non-Abelian gauge theory (``Technicolor'') and additional massless
fermions (``technifermions'') which feel this new force. The global
chiral symmetry of the fermions is spontaneously broken by the
formation of a technifermion condensate, just as the approximate
chiral $SU(2) \times SU(2)$ symmetry in QCD is broken down to $SU(2)$
isospin by the formation of a quark condensate. If the quantum numbers
of the technifermions are chosen correctly ({\it e.g.}\/ by choosing
technifermions in the fundamental representation of an SU$(N)$
Technicolor gauge group, with the left-handed technifermions being
weak doublets and the right-handed ones weak singlets) this condensate
can break the electroweak interactions down to electromagnetism.

The breaking of the global chiral symmetries implies the existence of
Goldstone bosons, the ``technipions'' ($\pi_T$).  Through the Higgs
mechanism, three of the Goldstone bosons become the longitudinal components
of the $W$ and $Z$, and the weak gauge bosons acquire a mass proportional to
the technipion decay constant (the analog of $f_\pi$ in QCD). The quantum
numbers and masses of any remaining technipions are model dependent. There
may be technipions which are colored (octets and triplets) as well as those
carrying electroweak quantum numbers, and some color-singlet technipions are
too light\cite{Eichten:1979ah,tpnumbers} unless additional sources of
chiral-symmetry breaking are introduced. The next lightest Technicolor
resonances are expected to be the analogs of the vector mesons in QCD. The
technivector mesons can also have color and electroweak quantum numbers and,
for a theory with a small number of technifermions, are expected to have a
mass in the TeV range\cite{Dimopoulos:1981yf}.

While Technicolor chiral symmetry breaking can give mass to the $W$
and $Z$ particles, additional interactions must be introduced to
produce the masses of the standard model fermions. The most thoroughly
studied mechanism for this invokes ``extended Technicolor'' (ETC)
gauge interactions\cite{Eichten:1979ah,Dimopoulos:1979es}. In ETC,
Technicolor, color and flavor are embedded into a larger gauge group
which is broken to Technicolor and color at an energy scale of
100s to 1000s of TeV.  The massive gauge bosons associated with this breaking
mediate transitions between quarks/leptons and technifermions, giving
rise to the couplings necessary to produce fermion masses.  The ETC
gauge bosons also mediate transitions among technifermions themselves,
leading to interactions which can explicitly break unwanted chiral
symmetries and raise the masses of any light technipions.  The ETC
interactions connecting technifermions to quarks/leptons also mediate
technipion decays to ordinary fermion pairs. Since these interactions
are responsible for fermion masses, one generally expects technipions
to decay to the heaviest fermions kinematically allowed (though this
need not hold in all models).

In addition to quark masses, ETC interactions must also give rise to
quark mixing. One expects, therefore, that there are ETC interactions
coupling quarks of the same charge from different generations. A
stringent limit on these flavor-changing neutral current interactions
comes from $K^0$--$\overline K^0$ mixing\cite{Eichten:1979ah}. These force
the scale of ETC breaking and the corresponding ETC gauge boson masses
to be in the 100-1000 TeV range (at least insofar as ETC
interactions of first two generations are concerned). To obtain quark
and technipion masses that are large enough then requires an
enhancement of the technifermion condensate over that expected naively
by scaling from QCD. Such an enhancement can occur if the Technicolor
gauge coupling runs very slowly, or ``walks''\cite{walking}. Many
technifermions typically are needed to make the TC coupling walk,
implying that the Technicolor scale and, in particular, the
technivector mesons may be much lighter than 
1~TeV\cite{Chivukula:1998if,Eichten:1996dx}.
It should also be noted that
there is no reliable calculation of electroweak parameters in a
walking Technicolor theory, and the values of precisely measured
electroweak quantities\cite{langacker} cannot directly be used to
constrain the models.

In existing colliders, technivector mesons are dominantly produced
when an off-shell standard model gauge-boson ``resonates'' into a
technivector meson with the same quantum numbers\cite{ehlq}. The
technivector mesons may then decay, in analogy with $\rho\to \pi\pi$,
to pairs of technipions.  However, in walking Technicolor the
technipion masses may be increased to the point that the decay of a
technirho to pairs of technipions is kinematically 
forbidden\cite{Eichten:1996dx}. In this case the decay to a technipion and a
longitudinally polarized weak boson (an ``eaten'' Goldstone boson) may
be preferred, and the technivector meson would be very narrow.
Alternatively, the technivector may also decay, in analogy with the
decay $\rho\to\pi \gamma$, to a technipion plus a photon, gluon, or
transversely polarized weak gauge boson.  Finally, in analogy with the
decay $\rho \to e^+ e^-$, the technivector meson may resonate back to
an off-shell gluon or electroweak gauge boson, leading to a decay into
a pair of leptons, quarks, or gluons.

\medskip
\subsection{Top Condensate and Related Models}

The top~quark is much heavier than other fermions and must be more strongly
coupled to the symmetry-breaking sector. It is natural to consider whether
some or all of electroweak-symmetry breaking is due to a condensate of
top~quarks\cite{Chivukula:1998if,topcondense}.  Top-quark condensation
alone, without additional fermions, seems to produce a top-quark mass
larger\cite{Bardeen:1990ds} than observed experimentally, and is therefore
not favored.  Topcolor assisted Technicolor\cite{Hill:1995hp} combines
Technicolor and top-condensation.  In addition to Technicolor, which provides
the bulk of electroweak symmetry breaking, top-condensation and the top quark
mass arise predominantly from ``topcolor,'' a new QCD-like interaction which
couples strongly to the third generation of quarks.  An additional, strong,
U(1) interaction (giving rise to a topcolor $Z'$) precludes the formation of
a $b$-quark condensate.

The top-quark seesaw model of electroweak symmetry
breaking\cite{Dobrescu:1998nm} is a variant of the original
top-condensate idea which reconciles top-condensation with a lighter
top-quark mass.  Such a model can easily be consistent with precision
electroweak tests, either because the spectrum includes a light
composite Higgs\cite{Chivukula:1998wd,Chivukula:2001er} or because
additional interactions allow for a heavier
Higgs\cite{Collins:1999rz,He:2001fz}. Such theories may arise
naturally from gauge fields propagating in compact extra spatial
dimensions\cite{dobrescu}.

A variant of topcolor-assisted Technicolor is flavor-universal, in
which the topcolor SU(3) gauge bosons, called colorons, couple equally
to all quarks\cite{equala,equalb}.  Flavor-universal versions of the
seesaw model\cite{colorons,universalseesaw} incorporating a gauged
flavor symmetry are also possible. In these models {\it all}\
left-handed quarks (and possibly leptons as well) participate in
electroweak symmetry-breaking condensates with separate (one for each
flavor) right-handed weak singlets, and the different fermion masses
arise by adjusting the parameters which control the mixing of each
fermion with the corresponding condensate.  A prediction of these
flavor-universal models, is the existence of new heavy gauge bosons,
coupling to color or flavor, at relatively low mass scales. A mass
limit of between 0.8 and 3.5~TeV is set\cite{Bertram:1998wf} depending
on the coloron-gluon mixing angle. Precision electroweak measurements
constrain\cite{Burdman:1999us} the masses of these new gauge bosons to
be greater than 1--3 TeV in a variety of models, for strong couplings.

\medskip
\subsection{Enhanced gauge-boson couplings and fermion compositeness}

If the new strong dynamics scale is somewhat higher than that accessible
to the next generation of colliders, the expected signature would be
enhanced gauge-boson self-interactions conventionally parameterized
by the ``anomalous couplings''\cite{anom:1987,dawson,herrero,wu:1993}, 
and the fermion contact interactions the so-called 
``fermion compositeness''\cite{comp} at a scale $\Lambda$.

Although the current LEP and Tevatron experiments have put stringent
bounds on the anomalous gauge-boson self-interactions, the anticipated
size of those couplings due to new strong dynamics may be of order
$v^2/\Lambda^2\sim 1/16\pi^2 < 10^{-3}$, smaller than the current
bounds. Experiments at future 
colliders will reach sensitivity to this level. In particular,
high energy scattering of longitudinal gauge-bosons $W_L,Z_L$ as
the electroweak Goldstone bosons should be the most direct probe
to the electroweak symmetry breaking sector. General arguments 
such as unitarity \cite{strong:1977,strong:1985} indicate that 
new physics associated with the electroweak symmetry breaking must
show up in some form at the scale of TeV, which can be
accessible most likely only at higher energy colliders 
of next generation. Regarding the
fermion compositeness, higher sensitivity will be reached
at higher energies due to the energy-dependent nature of
the dimension 6-operators \cite{comp}. In the next two
sections, we will summarize the studies of the
above physics scenarios at future colliders.

\section{Strong Dynamics at $e^+e^-$ Linear Colliders}

An $\ee$ linear collider with $\sqrt{s}=0.5-1.5$~TeV and a luminosity 
of $500-1000$ pb$^{-1}$ can be a very effective probe of strong electroweak
symmetry breaking.  Production mechanisms and
backgrounds are limited to electroweak processes, so that
signal and background cross sections can be  calculated exactly.
The initial state is well defined not only  in terms of four-momentum, 
but also 
in terms of electron (and possibly positron) helicity.
Also, complete final state helicity analyses are possible, 
due to the fact that most if not all of the  
final state kinematic variables can be reconstructed.

In this section we review the $\ee$ collider phenomenology of strong 
$\ww$ interactions which appear when there is no light Higgs particle 
with large couplings to vector gauge bosons. Detection of directly 
produced narrow-width spinless particles such as 
technipions \cite{Casalbuoni:1998fs} and top-pions \cite{Yue:2000xa}  
is straightforward up to the kinematic limit, 
and will not be discussed further.

\subsection{$\ee\ra\nu \bar{\nu}\ww$,\ \   
$\nu \bar{\nu}ZZ,\ \  \ww Z,\ \ ZZZ, \ \ \nu \bar{\nu} t\bar{t}$}

The first step in studying the reaction $\ee\ra\nu \bar{\nu}\ww$ is 
to separate the scattering of a pair of longitudinally polarized $W$'s, 
denoted by $\wwl$,
from transversely polarized $W$'s and background such as 
$\ee\ra\ee\ww$ and $e^- \bar{\nu}W^+Z$.
Studies have shown that simple cuts\cite{Barger:1995cn} 
can be used to achieve this separation 
in $\ee\ra\nu \bar{\nu}\ww$, $\nu \bar{\nu}ZZ$
at $\sqrt{s}=1000$~GeV, and that the signals\cite{Boos:1998gw,Boos:1999kj}
are comparable to those obtained 
at the LHC\cite{barger:1990,bagger:1994,Chanowitz:1994zh,bagger:1995}.
%
%
Furthermore, by analyzing the gauge boson production
and decay angles it is possible to use these reactions to 
measure chiral Lagrangian parameters with an accuracy greater 
than that which can be achieved at the LHC~\cite{Chierici:2001ar}.

The chiral Lagrangian parameters associated with quartic gauge boson couplings
can also be measured with the triple gauge boson production 
processes $\ee\ra\ww Z$ and 
$\ee\ra ZZZ$~\cite{Barger:1988sq,Barger:1989fd,Han:1998ht}.  
These measurements complement the  $\ww$ fusion measurements, 
and they will play a crucial role in 
multi-parameter chiral Lagrangian analyses.

The reaction $\ee\ra \nu \bar{\nu} t\bar{t}$ provides unique access to 
$\ww\ra t\bar t$ since this process is overwhelmed 
by the background $gg\ra t\bar{t}$ at the LHC.  
Techniques similar to those employed
to isolate $\wwl\ra \ww, ZZ$ can be used to measure
the enhancement in  $\wwl\ra t\bar{t}$ 
production\cite{Barklow:1996ti,Larios:2000xj,RuizMorales:1999kz,Han:2000ic}.
Even in the absence of a resonance it will be possible 
to establish a clear signal.  The ratio $S/\sqrt{B}$ is 
expected to be 12 for a linear collider with $\sqrt{s}=1$~TeV, 
1000 fb$^{-1}$ and $80\%/0\%$ electron/positron beam polarization,
increasing to 22 for the same luminosity and beam polarization 
at $\sqrt{s}=1.5$~TeV.

\subsection{$\ee\ra\ww$}

Strong gauge boson interactions induce anomalous triple gauge 
couplings (TGC's) at tree-level\cite{anom:1987,wu:1993,dawson,herrero}:
\begin{eqnarray}
   \kappa_\gamma &=& 1+\frac{e^2}{32\pi^2s_w^2}\bigl(L_{9L}+L_{9R}\bigr) 
\nonumber
 \\  \kappa_Z  &=& 1+\frac{e^2}{32\pi^2s_w^2}
   \bigl(L_{9L}-\frac{s_w^2}{c_w^2}L_{9R}\bigr) \nonumber
 \\  g_1^Z  &=& 1+\frac{e^2}{32\pi^2s_w^2}\frac{L_{9L}}{c_w^2} \  \nonumber .
\end{eqnarray}
where $\kappa_\gamma$, $ \kappa_Z$, and $g_1^Z$ are TGC's,
      $s_w^2=\sin^2\theta_w$,
      $c_w^2=\cos^2\theta_w$, 
and $L_{9L}$ and  $L_{9R}$ are chiral Lagrangian 
parameters\cite{Bagger:1993vu}.
Assuming QCD values for $L_{9L}$ and  $L_{9R}$, $\kappa_\gamma$
is shifted by  $\dkg \sim -3\times 10^{-3}$.
\begin{table}[]
\begin{center}
\begin{tabular}{l|cc|cc}
     & \multicolumn{4}{c}{error $\times 10^{-4}$} \\
\hline
     & \multicolumn{2}{c|}{$\sqrt{s}=500$ GeV} &  
\multicolumn{2}{c}{$\sqrt{s}=1000$ GeV} \\
 TGC & Re & Im & Re & Im \\
\hline
& & & & \\
$g^\gamma_1$ &  15.5    & 18.9     & 12.8     &  12.5     \\
$\kappa_\gamma$ &  \ 3.5    &  \ 9.8    & \ 1.2     &   \ 4.9    \\
$\lambda_\gamma$ & \ 5.4     &  \ 4.1    &  \ 2.0    &   \ 1.4    \\
$g^Z_1$          &  14.1    & 15.6     & 11.0     &  10.7     \\
$\kappa_Z$        & \ 3.8     &  \ 8.1    &  \ 1.4  &  \ 4.2    \\
$\lambda_Z$        &  \ 4.5     & \ 3.5     &  \ 1.7  &  \ 1.2  \\
\hline
\end{tabular}
\caption{Expected errors for the real and imaginary parts of 
CP-conserving TGCs assuming $\sqrt{s}=500$~GeV, 
${\cal L}=500$~fb$^{-1}$ and  $\sqrt{s}=1000$~GeV, 
${\cal L}=1000$~fb$^{-1}$.  The results are for
one-parameter fits in which all other TGCs are kept fixed at 
their SM values.}
\label{tab:cp-conserving}
\end{center}
\end{table}

Table~\ref{tab:cp-conserving} contains the estimates of the TGC
precision that can be obtained at $\sqrt{s}=500$ and 1000~GeV for the
CP-conserving couplings  $g^V_1$, $\kappa_V$, and 
$\lambda_V$~\cite{Abe:2001wn}.  These estimates are derived
from one-parameter fits in which all other TGC parameters are kept fixed 
at their tree-level SM values.  
The $4\times 10^{-4}$ precision for the TGCs $\kappa_\gamma$ and $\kappa_Z$ 
at $\sqrt{s}=500$~GeV 
can be interpreted as a precision of $0.26$ for the chiral Lagrangian 
parameters $L_{9L}$ and $L_{9R}$.
Assuming naive dimensional analysis\cite{Manohar:1984md} 
such a measurement 
would provide a $8\sigma$ ($5\sigma$) signal for $L_{9L}$ and $L_{9R}$
if the strong symmetry breaking energy scale were 3~TeV (4~TeV).

When $\ww$ scattering becomes strong
the amplitude for $\ee\ra\wwl$ develops a complex form factor $F_T$
in analogy with the pion form factor in 
$\ee\ra\pi^+\pi^-$\cite{Peskin:1984xw,Iddir:1990xn}.  
To evaluate the size of this effect the 
following expression for $F_T$ can be used:
\[
 F_T =
         \exp\bigl[{1\over \pi} \int_0^\infty
          ds'\delta(s',M_\rho,\Gamma_\rho)
          \{ {1\over s'-s-i\epsilon}-{1\over s'}\}
         \bigr]
\]
where
\[
\delta(s,M_\rho,\Gamma_\rho) = {1\over 96\pi} {s\over v^2}
+ {3\pi\over 8} \left[ \
\tanh (
{
s-M_\rho^2
\over
M_\rho\Gamma_\rho
}
)+1\right] \ .
\]
Here $M_\rho,\Gamma_\rho$ are the mass and width respectively of 
a vector resonance in $\wwl$ scattering. The term 
\[
\delta(s) = {1\over 96\pi} {s\over v^2}
\]
is the Low Energy Theorem (LET) amplitude for $\wwl$ scattering 
at energies below a resonance.  
Below the resonance, the real part of $F_T$ is proportional to  
$L_{9L}+L_{9R}$ and can therefore be
interpreted as a TGC.   
The imaginary part, however, is a distinct new effect.

\begin{figure}[tbh] 
\centerline{\includegraphics[angle=-90,clip=,width=9cm]
{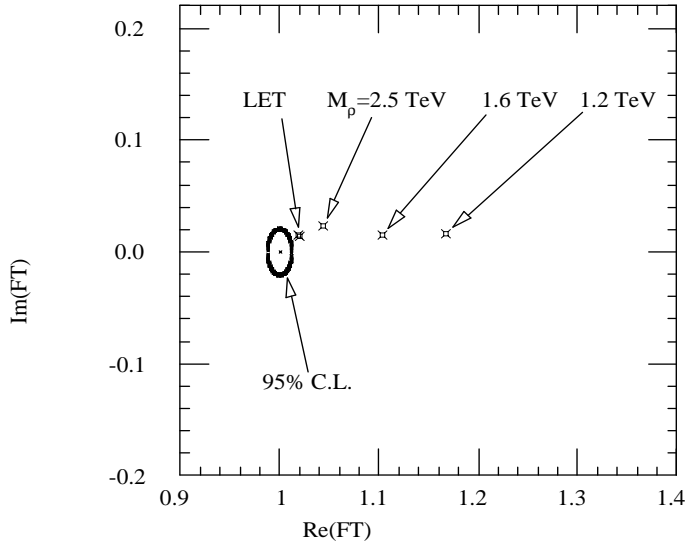}}
\vspace{10pt}
\caption{95\% C.L. contour for $F_T$  for $\sqrt{s}=500$~GeV
and 500~$fb^{-1}$. Values of $F_T$  for various masses $M_\rho$ 
of a vector resonance in $\wwl$ scattering
are also shown. The $F_T$ point ``LET'' refers to the
case where no vector resonance exists at any mass in strong $\wwl$ scattering.}
\label{fig:fteight}
\end{figure}

\begin{figure}[] 
\centerline{\includegraphics[height=17cm]
{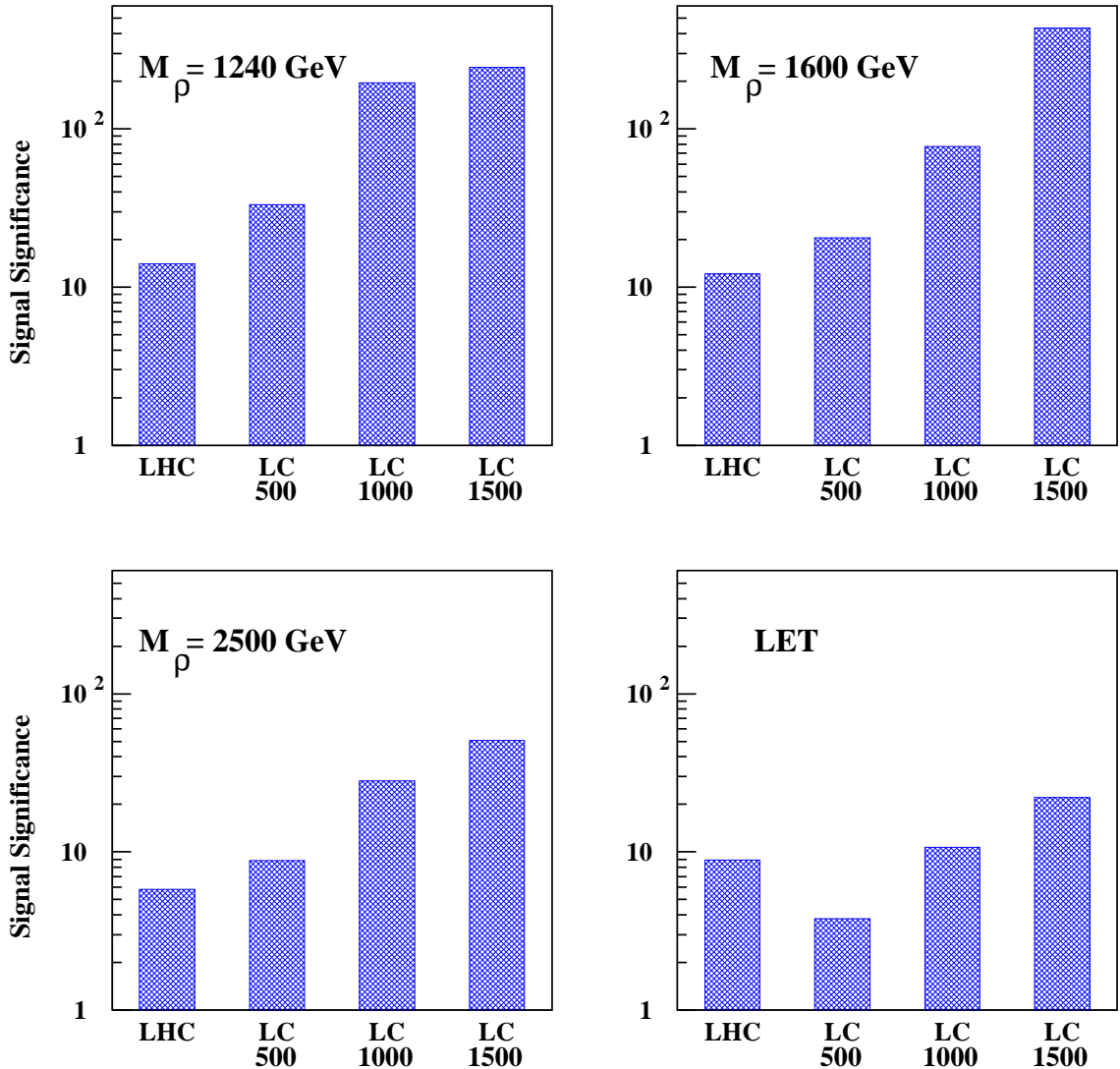}}
\vspace{10pt}
\caption{Direct strong symmetry breaking signal significance 
in $\sigma$'s for various masses $M_\rho$ of a vector resonance 
in $\wwl$ scattering.
The numbers below the ``LC'' labels refer to the
center-of-mass energy of the linear collider in GeV.
The luminosity of the LHC is assumed to be 300~$fb^{-1}$, 
while the luminosities
of the linear colliders are assumed to be  500, 1000, and 
1000~$fb^{-1}$ for 
$\sqrt{s}$=500, 1000, and 1500~GeV respectively.  
The lower right hand plot ``LET'' refers to the
case where no vector resonance exists at any mass 
in strong $\wwl$ scattering.}
\label{fig:strong_lc_lhc}
\end{figure}

The expected $95\%$ confidence level limits for $F_T$ for $\sqrt{s}=500$~GeV
and a luminosity of 500~$fb^{-1}$ are shown in Figure \ref{fig:fteight}, 
along with the predicted values of $F_T$ for various  masses $M_\rho$ of a 
vector resonance in $\wwl$ scattering.
The signal significances obtained by combining the results for 
$\ee\ra\nu \bar{\nu}\ww$, $\nu \bar{\nu}ZZ$\cite{Barger:1995cn,Boos:1998gw} 
with the $F_T$ analysis of $\ww$~\cite{Barklow:2000ci}
are displayed in Fig.~\ref{fig:strong_lc_lhc} along with the
results expected from the LHC\cite{unknown:1999fr}.   
At all values of the center-of-mass energy a linear collider
provides a larger direct strong symmetry breaking signal 
than the LHC for vector 
resonance masses of 1200, 1600 and 2500~GeV.   
Only when the vector resonance
disappears altogether (the LET case in the lower right-hand plot in 
Fig.~\ref{fig:strong_lc_lhc} ) does the  direct strong symmetry 
breaking signal from the $\sqrt{s}=500$~GeV linear collider
drop below the LHC signal. 
At higher $\ee$ center-of-mass energies the linear 
collider signal exceeds the LHC signal.

\section{Strong Dynamics at Hadron Colliders}

Hadron colliders offer exciting possibilities for searches for new particles 
and other signs of new strong dynamics and compositeness. High luminosity
$pp$ and $p\overline{p}$ machines should copiously produce proposed 
strongly-coupled resonances including technihadrons and excited quarks. They 
also probe contact interactions and vector boson scattering at extremely high 
energy scales. In this section we describe the expected physics 
reach of hadron 
colliders that exist (the Tevatron), are under construction (the LHC) and are
being designed (the VLHC).

\subsection{The Tevatron}

The Tevatron at Fermilab has taken approximately 100~pb$^{-1}$ of 
$p\overline{p}$ 
collision data at $\sqrt{s}=1.8$~TeV (Run I). In March 2001 Run II began, with
an increased energy ($\sqrt{s} = 1.96$~TeV) and a planned integrated luminosity
of 2~fb$^{-1}$ (Run IIa), followed by extended high luminosity running 
for a total in excess
of 15~fb$^{-1}$ per experiment. In Tables \ref{p1p3harrisa}--\ref{p1p3harrisd} 
announced 
results from Run I are tabulated along with extrapolations to RunIIa and a 
possible 30~fb$^{-1}$ complete RunII.

\begin{table}

\caption{\textbf{Sensitivity to Technicolor at the Tevatron}}
\label{p1p3harrisa}
\begin{tabular}{|rl|c|c|c|}
\hline\hline
\multicolumn{2}{|c|}{Channel} & Run I (100~pb$^{-1}$) & Run IIa 
(2~fb$^{-1}$) & Run II (30~fb$^{-1}$) \\
 &       & (GeV at 95\% CL)      & (GeV at 95\% CL)      & (GeV at 95\% CL) \\ 
\hline \hline
$\rho_{T1}\rightarrow W \pi_T$ & $\rightarrow l \nu b\overline{b}$ &
$170<M_\rho<200$\cite{p1p3h1} & $160<M_\rho<240$\cite{p1p3h2} & 
$M\rho<350-450$\cite{p1p3h3} \\
& &  (for $M_\pi\approx M\rho/2$) &  (for $M_\pi\approx M\rho/2$) & \\ 
\hline
$\omega_{T1}\rightarrow\gamma\pi_T$ & $\rightarrow\gamma b\overline{b}$ &
  $240<M_\omega<310$ ($M_\pi=120$) & - & - \\
& & $140<M_\omega<290$ ($M_\pi=60$)\cite{p1p3h4} & & \\ 
\hline
$\rho_{T1},\omega_{T1}$ & $\rightarrow e^+e^-$ &
$M<225$\cite{p1p3h5} & $M<410$\cite{p1p3h6} & - \\
\multicolumn{2}{|c|}{(If $W\pi$ and $\gamma\pi$ forbidden)} & & & \\
\hline
$\rho_{T8}\rightarrow qq,gg$ & $\rightarrow jj$ &
$260<M<480$\cite{p1p3h7} & $M<770$\cite{p1p3h9} & $M<900$\cite{p1p3h9} \\
($M_\pi>M_\rho/2$) & $\rightarrow b\overline{b}$ &
$350<M<440$\cite{p1p3h8} & & \\
\hline
$\rho_{T8}\rightarrow\pi_{LQ}\pi_{LQ}$ & $\rightarrow b\nu b\nu$ &
$M<600$\cite{p1p3h10} & $M<850$\cite{p1p3h12} & - \\
($M_\pi<M_\rho/2$) & $\rightarrow c\nu c\nu$ &
$M<510$\cite{p1p3h10} & - & - \\
& $\rightarrow b\tau b\tau$ &
$M<470$\cite{p1p3h11} & - & - \\
\hline\hline
\end{tabular}

\end{table}

\begin{table}
\caption{\textbf{Sensitivity to Topgluons at the Tevatron}}
\label{p1p3harrisb}

\begin{tabular}{|c|c|c|c|c|}
\hline\hline
Channel & Width & 
Run I (100~pb$^{-1}$) & Run IIa (2~fb$^{-1}$) & Run II (30~fb$^{-1}$) \\
                              &  $\Gamma/M$  & 
(GeV at 95\% CL)   & (GeV for $5\sigma$ signal)  & 
(GeV for $5\sigma$ signal) \\ 
\hline\hline
                                   & 0.3 & $280<M<670$  & $M<950$ & $M<1200$ \\
$g_T\rightarrow b\overline{b}$ & 0.5 & $340<M<640$\cite{p1p3h21} &
                                  $M<860$\cite{p1p3h22}  & 
$M<1100$\cite{p1p3h22} \\
                                   & 0.7 & $375<M<560$  & $M<770$ & $M<1000$ \\
 
\hline
                                   & 0.3 & - & $M<1110$ & $M<1400$ \\
$g_T\rightarrow t\overline{t}\rightarrow l\nu+jets$ & 0.5 & - & 
$M<1040$\cite{p1p3h23} &
                                  $M<1350$\cite{p1p3h23} \\
                                   & 0.7 & - & $M<970$  & $M<1290$ \\ \hline
                                   & 0.3 & - & $M<1000$ & $M<1200$ \\
$g_T\rightarrow t\overline{t}\rightarrow 6\ jets$   & 0.5 & - & 
$M<900$\cite{p1p3h24} &
                                  $M<1130$\cite{p1p3h24} \\
                                   & 0.7 & - & $M<800$  & $M<1100$ \\ 
\hline\hline
\end{tabular}

\end{table}

\begin{table}
\caption{\textbf{Sensitivity to Topcolor $Z'$ and $h_b$ at the Tevatron}}
\label{p1p3harrisc}

\begin{tabular}{|c|c|c|c|c|}
\hline\hline
Channel & Width & 
Run I (100~pb$^{-1}$) & Run IIa (2~fb$^{-1}$) & Run II (30~fb$^{-1}$) \\
                              &  $\Gamma/M$  & (GeV at $95\%$ CL)   & 
(GeV for $5\sigma$ signal)  & (GeV for $5\sigma$ signal) \\ 
\hline\hline
$Z'$ Model I\footnote{$Z'$ models described in \cite{p1p3h31} }
$\rightarrow t\overline{t}\rightarrow l\nu + jets$ & 0.02 & - & - & 
$M<830$\cite{p1p3h24} \\
                                                 & 0.04 & - & - & $M<670$\\ 
\hline
$Z'$ Model II $\rightarrow t\overline{t}\rightarrow l\nu + jets$ 
                             & 0.02 & - & $M<720$\cite{p1p3h24} & 
$M<980$\cite{p1p3h24} \\
                             & 0.04 & - & $M<950$               & $M<1200$ \\ 
\hline
$Z'$ Model III $\rightarrow t\overline{t}\rightarrow l\nu + jets$ 
                             & 0.02 & - & $M<600$\cite{p1p3h24} & 
$M<910$\cite{p1p3h24} \\
                             & 0.04 & - & $M<800$               & $M<1000$ \\ 
\hline
                      & 0.012 & $M<480$ & -                     & - \\
$Z'$ Model IV $\rightarrow t\overline{t}\rightarrow l\nu + jets$
        & 0.02  & $M<650$\cite{p1p3h32} & $M<980$\cite{p1p3h24} & 
$M<1200$\cite{p1p3h24} \\
        & 0.04  & $M<780$               & $M<1100$              & $M<1300$ \\ 
\hline
\multicolumn{3}{|c|}{} &  (GeV at 95\% CL)      & (GeV at 95\% CL) \\ \hline
$b\overline{b}h_b\rightarrow b\overline{b} b\overline{b}$ & & - & 
$M<270$\footnote{Using
$y_b/y_b^{SM}=72$ in Fig 8b of \cite{p1p3h34}} & $M<380$ \\ \hline\hline
\end{tabular}

\end{table}

\begin{table}
\caption{\textbf{Sensitivity to Compositeness at the Tevatron.} 
In each channel, $\Lambda^+$ is the upper entry and $\Lambda^-$ the lower.}
\label{p1p3harrisd}

\begin{tabular}{|c|c|c|c|}
\hline\hline
Channel & Run I (100~pb$^{-1}$)     & Run IIa (2~fb$^{-1}$) & Run II 
(30~fb$^{-1}$) \\
        & (TeV at 95\% CL)          & (TeV at 95\% CL)      & (TeV at 95\% CL) 
\\ \hline \hline
$\Lambda^\pm(qq\rightarrow qq)$      & 2.7\cite{p1p3h41} & - & - \\
                                    & 2.4               & - & - \\ \hline
$\Lambda^\pm(qq\rightarrow ee)$      & 3.3\cite{p1p3h42} & 6.5\cite{p1p3h43} & 
14\cite{p1p3h43} \\
                                    & 4.2               & 10                & 2
0 
   \\ \hline
$\Lambda^\pm(qq\rightarrow\mu\mu)$  & 2.9\cite{p1p3h44} & - & - \\
                                    & 4.2               & - & - \\ \hline
$\Lambda^\pm(qq\rightarrow\gamma\gamma)$ & - & 0.75\cite{p1p3h43,p1p3h45} 
                                              & 0.9\cite{p1p3h43,p1p3h45} 
\\
                                         & - & 0.71 & - \\ \hline
$q^*\rightarrow q\gamma,qW$         & 0.54\footnote{25~pb$^{-1}$}\cite{p1p3h46}
 
& 0.91\cite{p1p3h9} &
                                      1.18\cite{p1p3h9} \\ \hline
\multicolumn{2}{|c|}{} & (TeV for $5\sigma$ signal)  & (TeV for $5\sigma$ 
signal) \\ \hline
$q^*\rightarrow qg$  & 0.76\footnote{D\O\ $q^*$ search (Bertram) combined with 
\cite{p1p3h48}} 
& 0.94\cite{p1p3h49} & 1.1\cite{p1p3h49} \\ \hline\hline
\end{tabular}

\end{table}

\subsection{The LHC}

Despite the challenge at hadron colliders in the search for
new strong dynamics at the TeV scale, much theoretical work
has been performed at the 
LHC\cite{barger:1990,bagger:1994,Chanowitz:1994zh,bagger:1995}.
Many studies of strong EWSB at ATLAS and CMS have been performed and summarized
in several places\cite{atlastdr,p1p3ianh,p1p3yellow,New}. An expected 
``low luminosity'' period will collect 30~fb$^{-1}$ of data 
at $\sqrt{s}=14$~TeV over the first three years of operation, 
and will be followed by a similar ``high
luminosity'' period collecting up to 300~fb$^{-1}$. High luminosity running
(up to $10^{34}$~cm$^{-2}$s$^{-1}$) presents many experimental
challenges with an average of 20 collisions per beam crossing, 
degrading tracking
and electron identification capabilities particularly in the forward region.

As an example of a Technicolor resonance search, ATLAS have considered 
the production
of 500~GeV technirho in a multiscale Technicolor model and its signal in the 
channel $\rho_T^\pm\rightarrow WZ\rightarrow l^\pm\nu l^+l^-$ \cite{atlastdr}.
This study assumes the 30~fb$^{-1}$ of low luminosity data and hence the full
lepton ID and tracking capabilities of the detector.
The expected signal significance is strongly dependent on the input 
model parameters:
a narrow resonance ($\Gamma_{\rho_T} = 1.1$~GeV) which is not allowed to decay
to $\pi_T\pi_T$ ($m_{\pi_T}>m_{\rho_T}/2$) could have $S/\sqrt{B}\approx 80$;
but for $\Gamma_{\rho_T}=110$~GeV and $m_{\pi_T}=110$~GeV this would drop to
an indiscernable $S/\sqrt{B}\approx0.3$.

The masses of observable resonances at the LHC are expected to be 
5-10$\times$ those
at the Tevatron. A $Z'$ with couplings similar to those of the 
Standard Model $Z$ should
be observable up to $m_{Z'}\approx5$~TeV and direct observation of 
excited quarks of 
$m_{q^*}\approx 6$~TeV is possible\cite{p1p3h49}. The reach for 
compositeness scales
is similarly enhanced, with 300~fb$^{-1}$ of dijet data being sensitive to 
$\Lambda\approx 40$~TeV.

A further possibility at the LHC is that as $\sqrt{\hat{s}}$ begins to 
exceed 1~TeV, strong interaction effects in $WW$ scattering could 
become detectable. 
If jets can be reliably tagged in the forward region at high 
luminosities, a signal
should be observable with the full 300~fb$^{-1}$.

\subsection{The Super-LHC}

There has been some discussion of upgrading the LHC in luminosity and energy
after the 300~fb$^{-1}$ run is complete. A possible (though unlikely) doubling 
of the energy has been considered along with a tenfold increase in 
instantaneous
luminosity. Since the LHC detectors were not designed for these 
conditions only 
jet and muon information is likely to be usefull. Such an upgrade could double
the reach for a $Z'$ ($m_{Z'}\approx 10$~TeV) and compositeness 
($\Lambda\approx 80$~TeV), and significantly increase the sensitivity for
excited quarks ($m_{q^*}\approx 9$~TeV) and the scale of $WW$ scattering
available ($\sqrt{\hat{s}}\approx1.5$~TeV, assuming that forward jet
tagging is still possible).
Unfortunately, most of these gains
come from the energy increase, which is less plausible than a simple
luminosity upgrade.

\subsection{The VLHC}

A staged 40-175~TeV $p\overline{p}$ collider operating at luminosities 
comparable to the LHC (1-2$\times10^{34}$~cm$^{-2}$s$^{-1}$) has been 
proposed\cite{vlhctdr}. Studies of such a machine's physics reach are
in progress (see also the E4 Working Group report), but the direct
reach for excited quark resonances is expected to be 
$m_{q^*}\approx 25$~TeV for 10~fb$^{-1}$ at $\sqrt{s}=100$~TeV\cite{p1p3h49},
and $WW$ scattering could be probed at the scale of $2-3$~TeV.

\begin{figure}[] %
\centerline{\includegraphics[height=10cm]{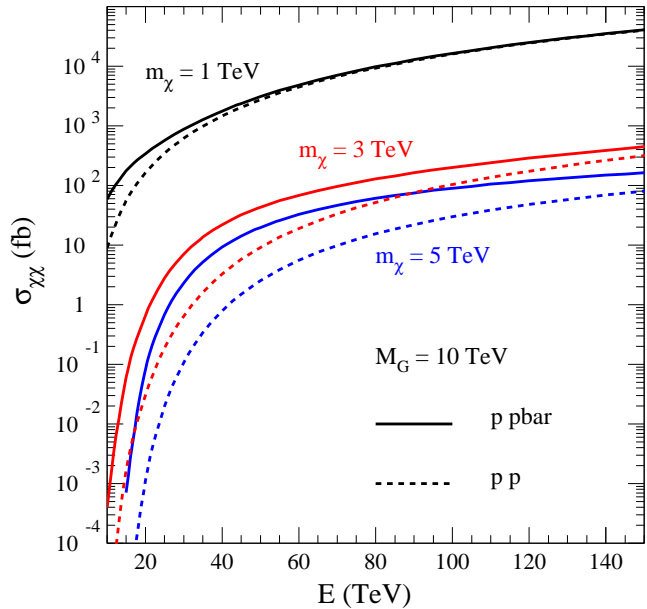}}
\vspace{10pt}
\caption{$\chi$-pair production in top-color models
at high energy hadron colliders leading to 6-top events.
}
\label{fig:timt}
\end{figure}

New signatures could become detectable at such high center-of-mass 
energies. For example,
in topcolor models, direct $\chi$ pair production and subsequent decays 
$\chi\rightarrow ht\rightarrow t\bar{t}t$ could occur\cite{He:2001fz}, 
with a $6t$ final state. 
Such a heavy state may only be copiously produced.
The
cross section for this process with $m_\chi=1$~TeV would be $\sim 10$~pb,
as shown in Fig.~{\ref{fig:timt}}.

\begin{figure}[] %
\centerline{\includegraphics[height=8.5cm]{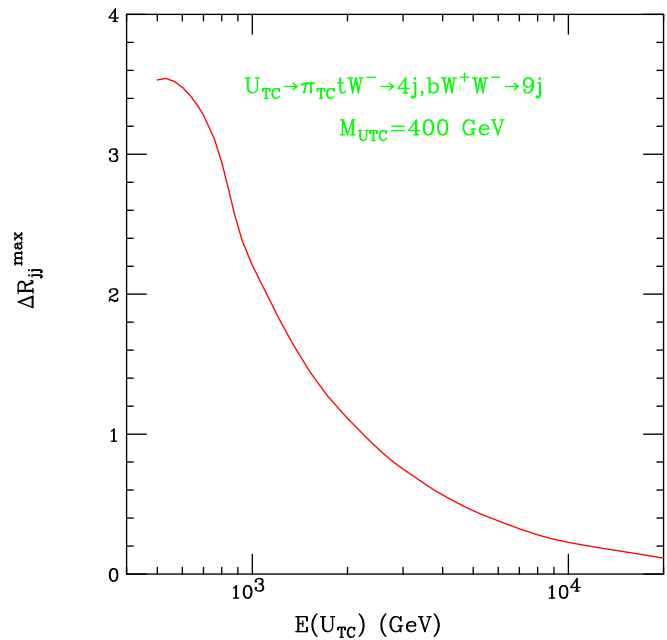}}
\vspace{10pt}
\caption{Maximum separation in $\Delta R$ distribution among
the jets from a heavy techni-quark decay.
}
\label{fig:tao}
\end{figure}

In interactions with $\sqrt{\hat{s}}\gg \Lambda_{TC}$, 
it is possible (in analogy
with QCD) that asymptotically free techniquarks could be produced 
that subsequently
hadronize into technijets consisting of weak vector bosons and 
technihadrons. A technijet
would manifest itself as an extremely massive but significantly boosted
(and hence not necessarily wide) jet in a VLHC detector. 
The production rate
for such a process can be significant:
For $m_{Q_T}=400$~GeV with $\sqrt{s}=100$~TeV the dijet 
differential cross section
for technijets exceeds that for $t\overline{t}$ for dijet masses 
$>900$ GeV. 
Exploration of technijets could provide the ultimate determination 
of the TC dynamics. As shown in Fig.~\ref{fig:tao}, a representative
techni-quark may decay subsequently into multiple jets and the 
separation between any two jets may be small enough so that 
experimental signature would be a very massive (but not too fat)
jet.

%
%

%
%

%
\begin{acknowledgments}
We would like to thank the participants in the Strong Electroweak 
Symmetry Breaking Working group for the contribution
during the Snowmass workshop, on which the materials summarized
in this report are based. 
T.L.B. was supported by Department of Energy contract DE-AC03-76SF00515.
R.S.C. was supported in part by a DOE grant DE-FG02-91ER40676.
T.H. was supported in part by
a DOE grant No.~DE-FG02-95ER40896 and in part by
the Wisconsin Alumni Research Foundation.
\end{acknowledgments}

\bibliography{p1p3_strong_dynamics}

\end{document}